\documentclass[iop, apj, tighten, onecolumn]{emulateapj}
\def\mag100{mag/100$^{\rm d}$}

\shorttitle{}

\shortauthors{McClelland et al.}

\begin{document}

\submitted{Submitted to ApJ, Jan 11, 2013}

\title{The Mid-Infrared and Optical Decay of SN~2011fe}

\author{
Colin M. M$^{\mathrm{c}}$Clelland\altaffilmark{1},
Peter M. Garnavich\altaffilmark{1},
Peter A. Milne\altaffilmark{2},
Benjamin J. Shappee\altaffilmark{3},
Richard W. Pogge\altaffilmark{3}
}
\altaffiltext{1}{Department of Physics, University of Notre Dame, Notre Dame IN 46556}
\altaffiltext{2}{Steward Observatory, University of Arizona, Tucson AZ 85719}
\altaffiltext{3}{Astronomy Department, The Ohio State University, Columbus OH 43210}

\begin{abstract}

We measure the decay rate of the mid-IR luminosity from type~Ia supernova 2011fe between six months and one year after explosion using Spitzer/IRAC observations. The fading in the 3.6~$\mu$m channel 
is 1.48$\pm 0.02$ \mag100, which is similar to that seen in blue optical bands. The supernova
brightness fades at 0.78$\pm 0.02$ \mag100 in the 4.5~$\mu$m channel which is close to that observed
in the near-IR. We argue that the difference is a result of doubly ionized iron-peak elements
dominating the bluer IRAC band while singly ionized species are controlling the longer wavelength
channel. To test this, we use Large Binocular Telescope spectra taken during the same phases
to show that doubly ionized emission lines do fade more slowly than their singly ionized
cousins. We also find that [\ion{Co}{3}] emission fades at more than twice the radioactive decay
rate due to the combination of decreasing excitation in the nebula, recombination and cobalt decaying to iron. The nebular emission velocities of [\ion{Fe}{3}] and [\ion{Co}{3}] lines show a smaller blue-shift
than emission from singly ionized atoms. The \ion{Si}{2} velocity gradient near maximum light
combined with our nebular velocity measurements suggest SN~2011fe was a typical
member of the `low velocity gradient' class of type~Ia.
Analyzing IRAC photometry from other supernovae we
find that mid-IR color of type~Ia events is correlated with the early light curve width and
can be used as an indicator of the radioactive nickel yield.
\end{abstract}


\section{Introduction}
Type-Ia supernovae (SNIa) are widely thought to be the thermonuclear explosions of degenerate white dwarfs (see \citealt{livio00} for a review).  Their high luminosity and uniformity make SNIa good cosmological distance indicators \citep{riess98,perlmutter99}, but a conclusive theoretical model of their explosion mechanism remains elusive.  For instance, it is an open question on whether the progenitor consists of a single degenerate white dwarf or two (or both). 

We do know that SNIa are bright because they synthesize a sizable
amount of radioactive $^{56}$Ni. Typically, SNIa produce half a Solar mass of radioactive
nickel which decays to cobalt ($\sim$1 week time scale) that then decays to stable $^{56}$Fe ($\sim$three month
time scale) emitting energetic $\gamma$-rays and positrons \citep{arnett82}. 
Early spectroscopy shows a significant amount of intermediate mass elements (IME) in the outer layers while
late-time spectra are dominated by singly and doubly ionized iron-peak elements \citep{filippenko97}.

SNIa explosions may begin with a subsonic burning front, creating a
deflagration of the carbon/oxygen white dwarf, but a deflagration alone yields
too little energy.  Alternatively, a pure detonation would create
too much radioactive yield and less IME than typically observed. 
A ``delayed detonation" (DD) better matches the observations. In the DD scenario \citep{khokhlov91}, fusion begins as a deflagration but switches to supersonic burning after some of the white dwarf has expanded, allowing carbon fusion at low density. 

Recent work has suggested that the dense deflagration region can be offset in position
and velocity from the larger, lower density detonation region \citep{maeda10}. This proposition
explains a number of observed SNIa properties near maximum and at later phases and that
some diversity in SNIa may simply be due to random viewing angles.

At late times, when the synthesized material becomes optically thin, the light from SNIa continues to be powered by radioactive decay. Months
to years after the explosion, the long-lived $^{56}$Co isotope keeps the nebula 
glowing by energy deposited from positrons while the $\gamma$-rays escape
without depositing energy.
The photometric decay is expected to be exponential with a e-folding time of 111.3 days
or 0.98 \mag100, matching that of $^{56}$Co decay. But \citet{lair06} found the
optical decline in the $B$ and $V$ bands is significantly faster than $^{56}$Co. In
contrast, \citet{sollerman04} showed the near-IR bands fade very slowly a
year after explosion. However, \citet{sollerman04} also demonstrated that the bolometric luminosity, or the total
flux across {\it all} bands from the optical to the IR, does decay at the expected $^{56}$Co energy deposition rate.

SN~2011fe \citep{nugent11} was the closest SNIa in 25 years ($\mu\approx29.04$; \citealt{shappee11}) and was extremely well
studied at early epochs.  It appears to have been a normal SNIa in terms of light curve,
luminosity and spectral development. 
Here, we present {\it Spitzer} IRAC photometry of SN~2011fe between six months and one year
after optical maximum. The brightness of SN~2011fe created a unique opportunity for detailed
study of a type~Ia event in the mid-IR, but it also appeared close to the
Earth's ecliptic pole, allowing several months of continuous Spitzer visibility.  To put these results into context, we also compare
{\it Spitzer} observations of SN~2011fe with data for other SNIa.
\label{s:introduction}

\section{Observations}
\label{s:observations}

\subsection{Spitzer/IRAC}

As part of a program to study the late-time properties of SNIa, we requested
low-impact target-of-opportunity observations of SN~2011fe with the {\it Spitzer} Space
Telescope using the Infrared Array Camera (hereafter IRAC; \citealt{fazio04}). IRAC
photometry was also obtained through Spitzer program 80196. The location of the supernova was
first visible to Spitzer at the end of January 2012 and remained visible until the following September.
Four visits were executed over the visibility period providing the best light curve of any SNIa
in the mid-IR. The supernova was observed in IRAC Channel~1 (CH1) and Channel~2
(CH2) which have effective wavelengths of 3.6$\mu$m and 4.5$\mu$m respectively.

Our program also obtained IRAC photometry of SN~2008Q and SN~2009ig, and we re-reduced archival IRAC data on SN~2005df previously presented by \citet{gerardy07}. These three SNIa were well-observed at optical wavelengths around maximum light 
\citep{mcclellandxx,foley12,milne10} and provide good comparisons to SN~2011fe.

We downloaded post-BCD images through the Spitzer Heritage Archive\footnote{http:/sha.ipac.caltech.edu/applications/Spitzer/SHA}.  Each SNIa
was identified in the IRAC images by its celestial coordinates\footnote{http://ned.ipac.caltech.edu/}.  
Aperture photometry was performed after the pixel values were converted to $\mu$Jy. For
SN~2011fe, we used the recommend 5-pixel aperture radius with annular sky radii of 12 to 20 pixels.  Pre-explosion {\it Spitzer} observations of host galaxy M101 conducted in 2004 provided background
estimates for both channels. 
The position of SN~2009ig in host galaxy NGC~1015 warranted a smaller 3-pixel aperture radius, while sky backgrounds were calculated via averaging the brightness fluctuations of multiple random apertures over areas without obvious point sources.  The same was done for SN~2008Q in host galaxy NGC~524 and SN~2005df in NGC~1559.  Aperture corrections were taken from the IRAC Instrument Handbook\footnote{http://irsa.ipac.caltech.edu/data/SPITZER/docs/irac/iracinstrumenthandbook/} 
and the results are presented in Table~\ref{t:photometry}.

\subsection{Large Binocular Telescope}

Spectra of SN~2011fe were obtained with the Large Binocular Telescope (LBT)
and Multi-Object Dual Spectrograph \citep[MODS;][]{pogge10} on several epochs during
its decline \citep{shappee12}. Three of the spectra, 2012 March 24, April 27 and June 12
(UT) were obtained over the span of the Spitzer observations and we use these to estimate the decay rates
of SN~2011fe at optical/near IR wavelengths. The epochs correspond to ages after $B$-band peak
light of 196, 230 and 276 days respectively.

The spectra were obtained through a 1.0 arcsecond wide slit that was oriented to the parallactic
angle for the March observation, taken at an airmass of 1.15. The April and June observations were
obtained at airmasses less than 1.1 so that differential slit losses were minimized.

The spectra were wavelength-calibrated using Neon and Argon emission line lamps taken during
the June run. The wavelength solution was then shifted based on the sky line centroids. The
MODS spectrograph is very stable and the wavelength shifts applied were less than 2~\AA.

MODS flux sensitivity is also very stable, so no flux calibration was applied. Instead, we directly
compared extracted electron count rates (proportional to the `analog-to-digital units')  to avoid further
uncertainties caused by flux calibrations using different standard stars under varying conditions.
To test the stability of the sensitivity, we divided MODS spectra of the standard star
HZ~44 obtained in April and June and found the fluxes differed by less than 5\%\ over the
spectral range between 3500~\AA\ and
8500~\AA. We therefore expect systematic uncertainty on the pixel by pixel decay rates to
by less than 0.05~\mag100. At wavelengths shorter than 3500~\AA\ and longer than 8500~\AA\ the difference between the standard star spectra reached 10\%, so the decay rates are not as reliable at the extreme ends of the spectra.

\begin{figure}[h!]
\epsscale{0.8}
\plotone{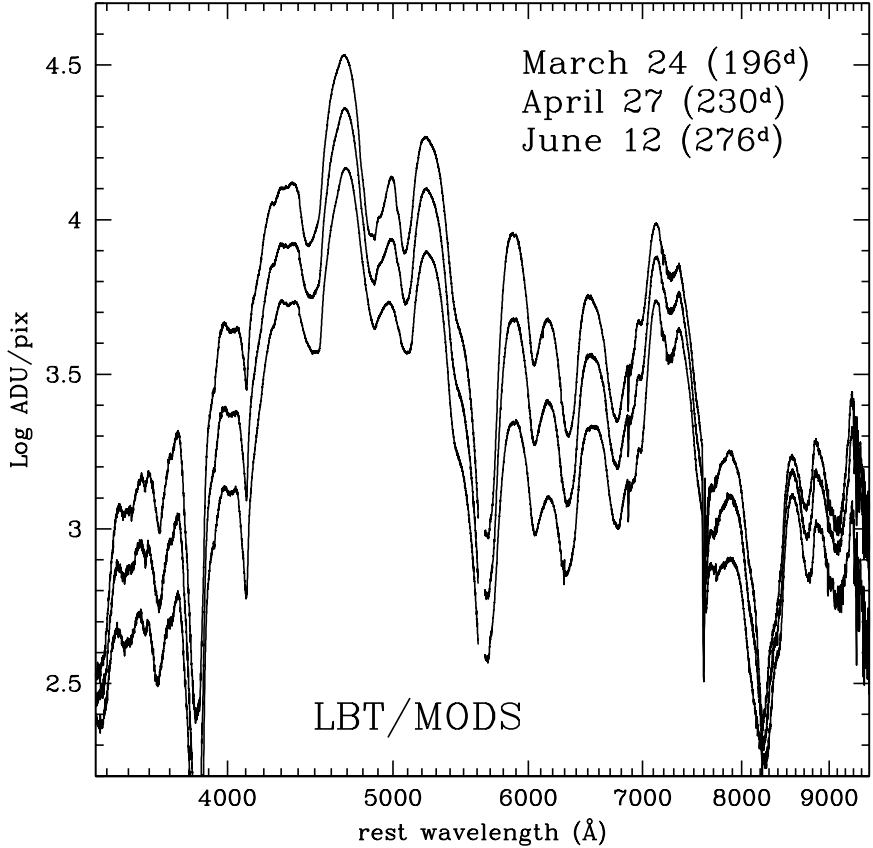}
\caption{\label{f:lbt} LBT/MODS spectra of SN~2011fe obtained during the 2012 Spitzer observations.
The spectra are not flux calibrated, but are plotted as the logarithm of the raw counts per pixel.
The MODS dichroic divides the red and blue sides of the spectrograph at 5700~\AA\ and this is
shown as a gap in the plot. Some emission features are seen to fade at a different rate than
others. For example, the emission peaks around 6000~\AA\ decline significantly faster than
the features at 7200~\AA.}
\end{figure}

%%%FIGURE LBT
The MODS spectra of SN~2011fe are shown in Figure~\ref{f:lbt}. In general, the spectra are
typical of a type~Ia supernova a couple hundred days after explosion. Between 4000~\AA\ and
5500~\AA\ there is a clump of emission dominated by forbidden lines of singly and doubly
ionized iron. Weaker lines from cobalt, iron and nickel are clearly seen long ward of 5500~\AA.
These are some of the highest signal-to-noise (S/N) spectra of a type~Ia supernova at
late phases ever obtained and they already been used to search for hydrogen emission from
SN~2011fe \citep{shappee12}.

The relative count rates across the spectra are stable from epoch to epoch, but the total efficiency
depends on seeing and cloud cover and these were not well controlled. To simulate a broadband filter, we multiplied the spectra
with a Bessell $B$-band filter function and integrated the result. 
From these values we adjusted the relative ADU to match the $B$-band decay rate of $1.32\pm0.02$ \mag100 found from the late-time photometry of \citet{munari13}.

LBT spectra of SN~2011fe were also obtained on 2012 May 1 (UT, age of 234 days) with the LUCIFER infrared spectrograph. The spectra covered the $J$ and $H$ bands using three grating settings.
Only one epoch of near-IR spectra was obtained so the decay rates at these wavelengths could
not be estimated, but the velocity offsets of the two strong [\ion{Fe}{2}] lines are measured.

\section{Analysis}
\label{s:analysis}
\begin{figure}[h!]
\epsscale{0.6}
\plotone{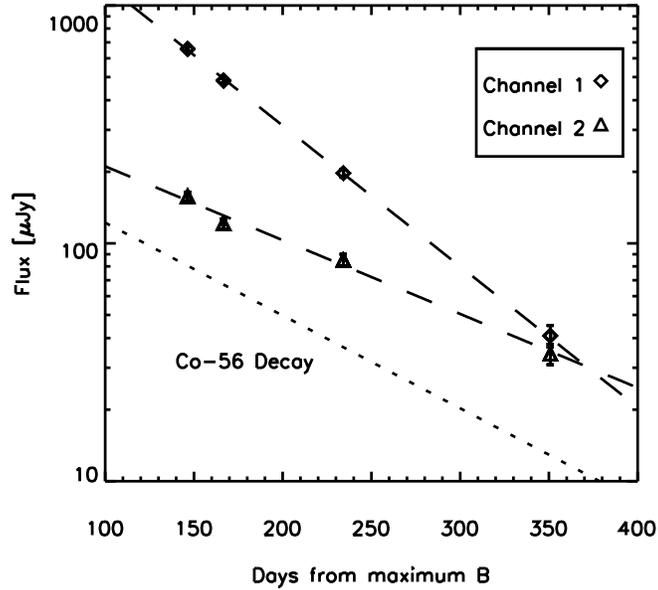}
\caption{\label{f:ir_lightcurve} 2011fe IR light curve computed from IRAC photometry.  The Channel~1 decay slope is consistent with typical SN $B$ and $V$ decay slopes, while Channel~2 decays faster than the $^{56}$Co energy deposition rate.  If the decay rate in each filter remains constant, 2011fe should see equal brightness in both channels at $\sim1$ year after maximum.}
\end{figure}

%%%%%%%FIGURE LIGHTCURVE
The mid-IR photometry of SN~2011fe starting at an age of five months and ending about one year after explosion is shown in Figure~\ref{f:ir_lightcurve}. The light curves are consistent with an exponential
decay, so we fit them with a function of the form $\mathcal{F}=A_{230}e^{-(t-230)/\tau}$, where
$\mathcal{F}$ is
the observed time varying flux, $t$ is the age since $B$ maximum, $A_{230}$ is the flux at the fiducial time of 230~days, and $\tau$ is the $e$-folding time of the decay. The decay rate can also
be described in magnitudes per 100~days by $\Delta = 108.57/\tau$.

The resulting measured decay rates for SN~2011fe are given in Table~\ref{t:irslopes}. The
1.48$\pm 0.02$ \mag100\ seen in CH1 is significantly steeper than the 0.78$\pm 0.2$ \mag100\ 
measured for CH2.  The decay observed in the two Spitzer bands straddle the $^{56}$Co rate of 0.98 \mag100. The decay rate for CH1 is similar to that typically seen in $B$ and $V$ bands,
while CH2 has a decay even more shallow than the $I$ band ($0.99\pm0.04$ \mag100; via least-squares fit to photometry from \citealt{munari13}).

The long cadence and greater distance to the other Spitzer-observed type~Ia events resulted in larger uncertainty in the measured decay rates compared with the SN~2011fe data.  The CH1 decay rates for SN~2008Q, SN~2009ig and SN~2005df all appear slightly slower. All the supernovae are consistent in that the CH1 rate is
steeper than the CH2 decline rate.

We created a color index based on the flux ratio between the two IRAC channels, or
$$\mathrm{CH1}-\mathrm{CH2} = -2.5 \log\left(\frac{\mathcal{F}_\mathrm{CH1}}{\mathcal{F}_\mathrm{CH2}}\right)+(ZP_\mathrm{CH1}-ZP_\mathrm{CH2})\,\,\,\,\mathrm{mag},$$
where $\mathcal{F}_\mathrm{CH1, CH2}$ are the fluxes measured in $\mu$Jy in CH1 and CH2, and $ZP_\mathrm{CH1,CH2}$ are the 
magnitude zero points for those passbands, listed in the IRAC Instrument Handbook.  
Extinction differences between these filters is assumed to be negligible \citep{flaherty07}.
Because the IRAC decay rates are so different, the CH1-CH2 color varies rapidly in time so we have chosen a fiducial time
of 230~days to estimate the color parameter. This age was chosen because the Spitzer archive
contains several SNIa observed near this age, thus reducing any errors caused by interpolation.

The estimated color at 230 days shows a wide range among the supernovae in our
sample. To better understand the cause of this diversity, we sought to compare the 230~day
color with early-time light curve shape, namely the parameter $\Delta m_{15}(B)$\footnote{$\Delta m_{15}(B)$ \citep{phillips93} is defined as the change in $B$ magnitude from maximum to 15 days post maximum.  It is related to the peak luminosity of a SN~Ia through the Phillips Relation.}.  For SN~2009ig we
use the $\Delta m_{15}(B)$ measured by \citet{foley12} and for SN~2011fe we
use the value from \citet{vinko12}. 
Both SN~200Qdf and 2008Q have excellent early light curves \citep{milne10,mcclellandxx}
and we measure their $\Delta m_{15}(B)$ directly. The $\Delta m_{15}(B)$ values and the
measured IRAC colors are given in Table~\ref{t:irslopes}.
 
\begin{figure}[h!]
\epsscale{0.5}
\plotone{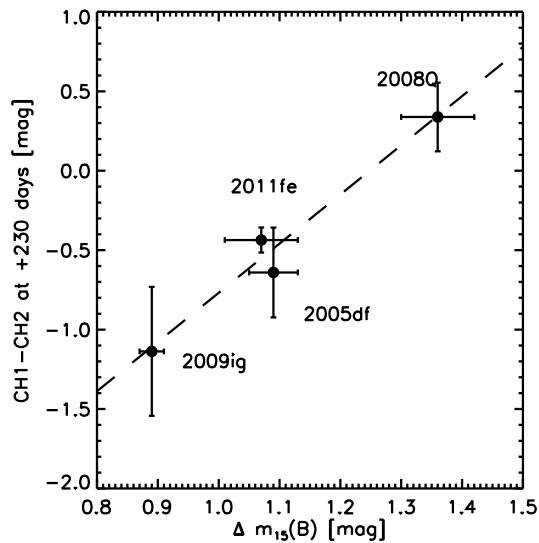}
\caption{\label{f:spitzer_color_dm15} Relation between late-time IRAC color and early-time decline rate.  Magnitudes are calculated according to the photometric zero points given in the IRAC Instrument Handbook.  The dashed line represents the least-squares linear fit to this relation.}
\end{figure}

 %%%%%%%FIGURE COLOR
The four supernovae, 2005df, 2008Q, 2009ig and 2011fe, show a wide range of early light curve decline rates, and so, a wide range of synthesized radioactive nickel masses. We plot our measured
IRAC colors at 230 days versus
$\Delta m_{15}(B)$ in Figure~\ref{f:spitzer_color_dm15}. There appears to be a robust correlation
(albeit with only four data points) indicating that the CH1 to CH2 flux ratio decreases
with increasing $\Delta m_{15}(B)$ parameter. Alternatively, we see that the CH1 flux increases
relative to CH2 as the amount of synthesized radioactive nickel increases since $\Delta m_{15}(B)$
correlates with nickel yield. Based on this
handful of supernovae, the IRAC color appears to be a good indicator of peak optical luminosity,
although this is unlikely to be useful for cosmological measurements.

\section{Discussion}
\label{s:discussion}

\subsection{IRAC Decay Rates}

The radically different late-time decay rates observed at 3.6$\mu$m and 4.5$\mu$m is puzzling
given that total flux from the optical to near-IR appears to match the $^{56}$Co energy injection
rate \citep{sollerman04, leloudas09}. The effect is also seen when comparing the faster decay in
the $B$ and $V$ bands to the modest $I$-band decay rate between six months and a year
after explosion. The fast optical decline is likely compensated by a slow near-IR decay
which after a year becomes a spectacular rise in the importance of the $J$ and $H$ band fluxes.

What do the fast declining wavelength bands ($B$, $V$, CH1) have in common and
how are they different from the slower ($I$, $J$, $H$, CH2) fading bands? As most of
the SNIa spectrum at late-times, the IRAC bands are expected to be dominated by emission
from iron-peak elements with a mixture of singly-ionized and doubly ionized species. \citet{gerardy07} predicted that  spectrum covered by CH1 would show strong [\ion{Fe}{3}] 3.229 $\mu$m
and [\ion{Co}{3}] 3.492 $\mu$m. In contrast, \citet{maeda10} predicts several strong
[\ion{Fe}{2}] should dominate the wavelength region spanned by CH2. 

We propose that the difference in the observed IRAC decay rates is a result of doubly ionized
species being the dominate light source in CH1 and singly ionized iron-peak species supplying the
majority of flux in CH2. 
The relative amounts of emission from singly and doubly ionized species should also vary in
the shorter wavelength bands, and we note that
the strong [\ion{Fe}{3}] 4700\AA\ blend straddles the $B$ and $V$ band transition
and contributes flux to both fast-decaying bands. In contrast, the slow-fading $I$, $J$ and $H$ bands
are predominately fed by singly ionized atoms. We predict the $K$-band around 2.2$\mu$m
should fade like IRAC CH1 since it contains a strong [\ion{Fe}{3}] line at 2.218$\mu$m.

Singly and doubly ionized iron-peak elements exist in very different regions of SNIa
nebulae. Using the W7 prescription \citep{thielemann86}, one-dimensional SNIa nebular models agree that [\ion{Fe}{2}]\ (and [\ion{Co}{2}] which has a similar ionization energy) atoms
dominate in the inner dense regions while doubly ionized iron occupies the outer volume
\citep[e.g.][]{mazzali11, liu97}. Generally, the dense regions cool quickly and
doubly ionized species recombine while cooling in low density regions is inefficient and
they remain dominated by doubly ionized gas. \citet{sollerman04} however, notes that photoionization
has a major effect on the ionization structure and it is not well modeled in the nebular spectra
\citep{kozma98}. 

\citet{sollerman04} modeled SNIa nebular spectra with and without accounting for photoionization. 
Without including photoionization in the calculation,
the [\ion{Fe}{3}]/[\ion{Fe}{2}] ratio was too small compared with the real spectrum of SN~2000cx at one year old. 
When including photoionization, the model predicted too much [\ion{Fe}{3}] which suggests
that real supernovae are intermediate between these two extremes. It is possible that the
effects of photoionization in SNIa nebulae decrease with time as the expansion reduces the
density. This should reduce the importance of the doubly ionized species through recombination while
increasing the flux of singly ionized species. As the nebula transitions away from
significant photoionization, we expect the filters dominated by lines from doubly ionized
species will fade more quickly than $^{56}$Co. Meanwhile, the bands dominated by singly
ionized atoms will fade more slowly than $\Delta = 0.98$ \mag100\ with the difference
in decay rates a clue to the recombination rate.

If the CH1 decay rate is a result of both the excitation decreasing at the $^{56}$Co rate and the
recombination rate from [\ion{Fe}{3}] $\rightarrow$ [\ion{Fe}{2}], then we can approximate it by 
$$exp(-t/\tau_\mathrm{CH1})=exp(-t/\tau_\mathrm{Co})\cdot exp(-t/\tau_\mathrm{rec}),$$
where $\tau_\mathrm{CH1}$, $\tau_\mathrm{Co}$ and $\tau_\mathrm{rec}$ are respectively the $e$-folding times of CH1, $^{56}$Co and [\ion{Fe}{3}] $\rightarrow$ [\ion{Fe}{2}] recombination.  Assuming collisional
ionization are negligible at late times and photoionization is becoming unimportant, the recombination should go approximately as $exp(-n_\mathrm{e}\alpha_R t)$, where $n_\mathrm{e}$ is the
electron number density and $\alpha_R$ is the recombination rate coefficient.  
The recombination coefficient is $\alpha_R\approx3\times10^{-12}$ cm$^{3}$ s$^{-1}$ and is nearly
constant for temperatures between $10^3$ and $10^6$~$^\circ$K \citep{nahar97}. We can
solve for the electron density, $n_\mathrm{e}=1/\alpha_R[1/\tau_\mathrm{CH1}-1/\tau_\mathrm{Co}]$
and find $n_\mathrm{e}=1.8\times 10^4$ cm$^{-3}$. Assuming most of the iron peak atoms are
doubly ionized, the mass density corresponding to this recombination rate is then approximately 
$10^{-18}$ g~cm$^{-3}$. 

We can compare this recombination density to the density of the standard W7 model scaled to
250 days after explosion \citep{mazzali11}. The scaled W7 density ranges between 
10$^{-16}$ g~cm$^{-3}$ in the inner regions to 10$^{-17}$ g~cm$^{-3}$ out at 10000~km~s$^{-1}$.
Considering the assumptions made in the calculation, this is surprisingly
close to the density for a pure recombination process removing doubly ionized species
from the nebula.

\subsection{Optical Decay Rates from LBT Spectra}
\begin{figure}[h!]
\epsscale{0.8}
\plotone{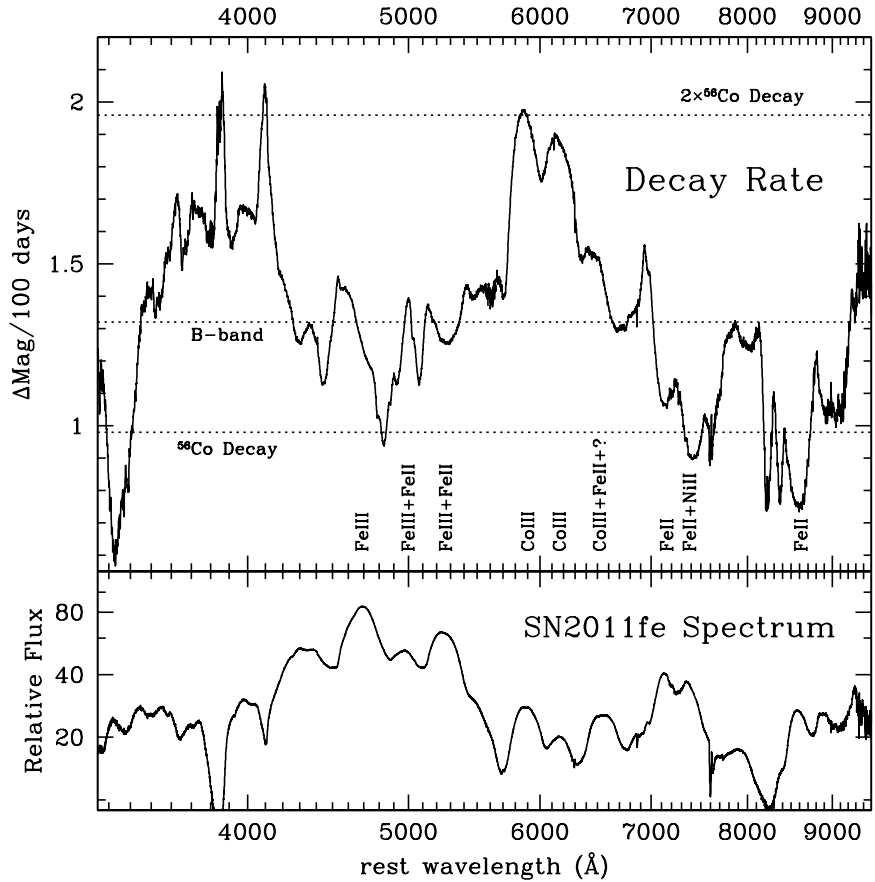}
\caption{\label{f:ratio} Decline rate $\Delta$ in units of \mag100 from the LBT/MODS spectra from 190 days to 270 days past maximum.  The optical spectrum of 2011fe is included for reference.  The horizontal dotted line at 1.41 \mag100 represents the calculated $B$-band decline rate, while the horizontal line at 0.98 \mag100 is the energy deposition rate, equal to $^{56}$Co decay.  We note that the large fading rate seen from 5700 \AA\ to 6400 \AA\ is \ion{Co}{3} which
has an additional loss channel through radioactive decay to iron.}
\end{figure}

%%%%%%%FIGURE RATIO
The ratio of the June~12 and March~24 count rates as a function of wavelength is shown in Figure~\ref{f:ratio} in units of \mag100. The ratio has been scaled to match the $B$-band
decay rate discussed in Section~\ref{s:observations}. Clearly, the decay rates vary by spectral region and by specific feature. Some of
the small-scale variation is due to shifts in emission centroids with time. For example, the
bright blend of [\ion{Fe}{3}] lines at 4700~\AA\ shifts by about 10~\AA\ redward over the 80~days between
spectra. This is shown as a decay rate of 1.4~\mag100\ on the blue side
of the feature and a 1.1~\mag100\ decay on the red side.  This drift was noted by \citet{maeda10} and appears typical in SNIa.
The total flux of the feature fades by 1.41~\mag100\ which is not surprising given that we have scaled the $B$-band flux to
1.32~\mag100\ and the 4700~\AA\ line contributed significantly to this filter.

In contrast, the weak features around 6000~\AA\ faded at twice the $^{56}$Co decay rate
and show no shift in the line centroid. The 5900 and 6160~\AA\ lines have been attributed
to [\ion{Co}{3}] emission, but the 5900~\AA\ line may also contain some Na~I D flux.

As noted in Figure~\ref{f:lbt}, the blend of bright lines around 7200~\AA\ fade rather
slowly compared with the [\ion{Fe}{3}] and [\ion{Co}{3}] lines in the blue. The emission around 7200~\AA\
is often attributed to [\ion{Fe}{2}] and [\ion{Ni}{2}] lines. The line at 8600~\AA\ also shows a slow
decay rate and it too has been identified as an [\ion{Fe}{2}] feature.

\begin{figure}[h!]
\epsscale{0.6}
\plotone{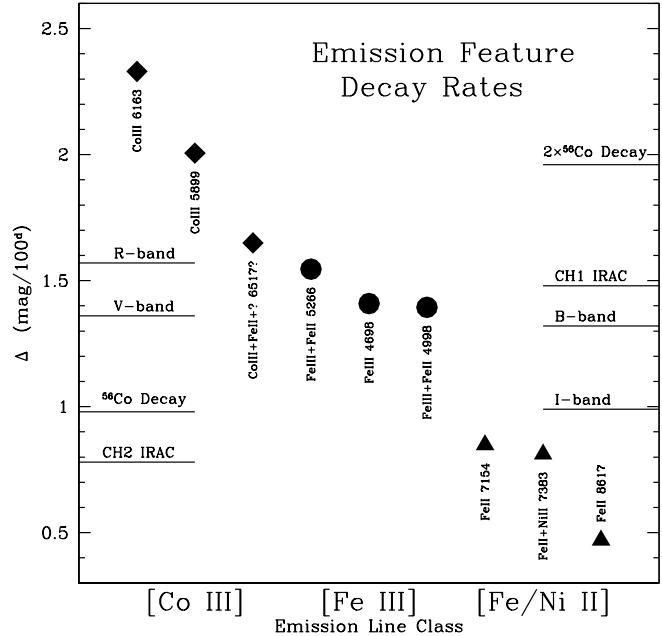}
\caption{\label{f:lines} Decay rate of the total flux for individual emission lines, measured between 24 March, 2012 and 12 June, 2012.  The approximate $BVRI$ \citep{munari13}, CH1 and CH2 decay rates of SN~2011fe are given for reference.}
\end{figure}

%%%%%%%FIGURE LINES
In general, the singly ionized species show the slow decay rates while emission from
doubly ionized atoms decays significantly faster than the radioactive decay rate
(see Figure~\ref{f:lines}). This
supports our interpretation of the decay rates we observed in the Spitzer bands and
that recombination is the likely sink for the doubly ionized species.

[\ion{Co}{3}] emission is hit with a triple whammy. The overall energy available for line
emission decreases at the $^{56}$Co radioactive decay rate. In addition,
recombination of doubly ionized atoms reduces their numbers as demonstrated in the
decay rate of [\ion{Fe}{3}] lines. But much of the cobalt is radioactive, so its numbers are
decreased at twice the radioactive decay rate plus the recombination loss rate.  

The emission at 6550~\AA\ has been identified as from [\ion{Co}{3}], but it appears to
fade at a slower rate the the other [\ion{Co}{3}] lines. 
The very strong shift in peak wavelength
as the line fades suggests it is a blend with a singly ionized emission feature and that
combination makes it appear to fade more slowly than the other [\ion{Co}{3}] lines.  
The NIST Atomic Spectra Database\footnote{http://physics.nist.gov/PhysRefData/ASD/lines\_form.html} points to strong [\ion{Fe}{2}] emission at 6446, 6456 and 6516 \AA, while \citet{liu97} suggests a combination of [\ion{Fe}{2}] and [\ion{Ni}{2}] in addition to [\ion{Co}{3}].
This feature's profile is flat-topped in the June spectrum and this makes it difficult to obtain
a precise peak velocity.

\subsection{IRAC Color Variation in Type Ia Supernovae}

Our supposition that emission from doubly ionized species decays at a faster rate
than those from singly ionized atoms appears to be supported by the optical spectra. 
Since IRAC CH1 shows a rapid decay rate, then doubly ionized species
should be dominant around 3.6$~\mu$m. Indeed, \citet{gerardy07} suggested
\ion{Fe}{3}\ and \ion{Co}{3}\ lines contribute significantly to the CH1 band.
Thus, the CH1$-$CH2 color observed in the SNIa might be a good indicator of the
ionization ratio within a SNIa nebula.

We can ask if the correlation between the IRAC color and $\Delta m_{15}(B)$ seen in
Figure~\ref{f:spitzer_color_dm15} implies that a [\ion{Fe}{3}]/[\ion{Fe}{2}] ratio
correlates with nickel yield.
SN~2003hv possessed notably strong [\ion{Fe}{3}] 4700 \AA\ at late times \citep{mazzali11}, though the SN had a large $\Delta m_{15}(B)$ of 1.6 \citep{leloudas09}, so probably synthesized a 
relatively small amount of radioactive nickel. This trend is the opposite of that implied by [\ion{Fe}{3}]
dominating the CH1 bandpass. \citet{mazzali11} argued that the very strong [\ion{Fe}{3}]
emission seen in SN~2003hv might be due to an unusually low nebular density and that the explosion
might be peculiar. The [\ion{Fe}{3}]/[\ion{Fe}{2}] ratio is probably a good indicator
of nebular density because the recombination rate is so sensitive to density. But it is
not clear why the nebular density at a fixed age should be related to $\Delta m_{15}(B)$.

If the fast decaying [\ion{Co}{3}] 3.492 $\mu$m line dominates the CH1 bandpass with contributions
from slower decaying emission such as [\ion{Fe}{2}] and [\ion{Fe}{3}], then the flux
would appear to decay at an intermediate rate over our Spitzer observation window. 
Such a scenario would explain Figure~\ref{f:spitzer_color_dm15} because the CH1$-$CH2 color
would correlate with the [\ion{Co}{3}]/[\ion{Fe}{2}] ratio and that
should be a good indicator of the radioactive nickel yield. This supposition awaits $L$-band spectra to identify the emission features in the CH1 bandpass.  

We note that the IRAC color could be matched in the optical using the SDSS $r-i$ color index
for SNIa at late times. The SDSS-$r$ band encompasses fairly uncontaminated
\ion{Co}{3} features while SDSS-$i$ band mainly sees emission by \ion{Fe}{2}. Such a
color at a fixed age should correlate with peak luminosity.

\subsection{Nebular Velocity}

\citet{maeda10} suggested that asymmetry in SNIa explosions could be manifested in the nebular spectra.
While most SNIa appear to show little asymmetry in the outer layers of the ejecta \citep{wang96}, \citet{maeda10b} showed that differences in velocity shifts between ions associated with dense, inner layers could be quite large when compared to velocity shifts of the outer layers.  The theory they develop is that of a dense, inner region rich in deflagration products that is offset in position and velocity
from the low-density, symmetric outer region.
In this scenario, doubly ionized iron-group lines tend to trace outer low-density regions, while their singly-ionized variants trace the dense inner regions.

We measured the wavelength centroids of several major features of the LBT/MODS spectra and found two distinct groups: [\ion{Fe}{3}] and [\ion{Co}{3}] lines tended to be blueshifted by $\sim$400 km s$^{-1}$, while [\ion{Fe}{2}] and [\ion{Ni}{2}] features were blueshifted between 1000 and 1500 km s$^{-1}$.  A list of these measured lines is given in Table~\ref{t:lineshift}.  Features made from blends of
singly and doubly ionized lines (e.g. the $\sim$5000 and $\sim$5250 \AA\ [\ion{Fe}{2}] + [\ion{Fe}{3}]) appear to show higher blueshifts, as well, suggesting that their singly-ionized components are dominant.  The highly blended $\sim$6550 feature proved difficult to model in this manner; the lack of a narrow, discernible peak meant a large error in identifying individual lines, though \citet{bowers97} identified [\ion{Co}{3}] 6578 \AA\ as an
important component, and the NIST Atomic Spectra Database points to major [\ion{Fe}{2}] contributions here, as well.  Synthetic spectra and improved atomic transition information
is necessary to better determine the shift of individual lines.

Despite the problems of line blending in the optical, it is clear that the velocity shifts of
the [\ion{Fe}{3}] 4700 \AA\ line and the [\ion{Co}{3}] features at $\sim$5900 and $\sim$6200 \AA\ 
are significantly smaller than the offsets of more than $-1000$~km$\;$s$^{-1}$ observed
in the singly ionized emission lines. Both [\ion{Fe}{3}]
and [\ion{Co}{3}] emission likely originate in the the outer, symmetric regions of the nebula. 

\begin{figure}[h!]
\epsscale{0.6}
\plotone{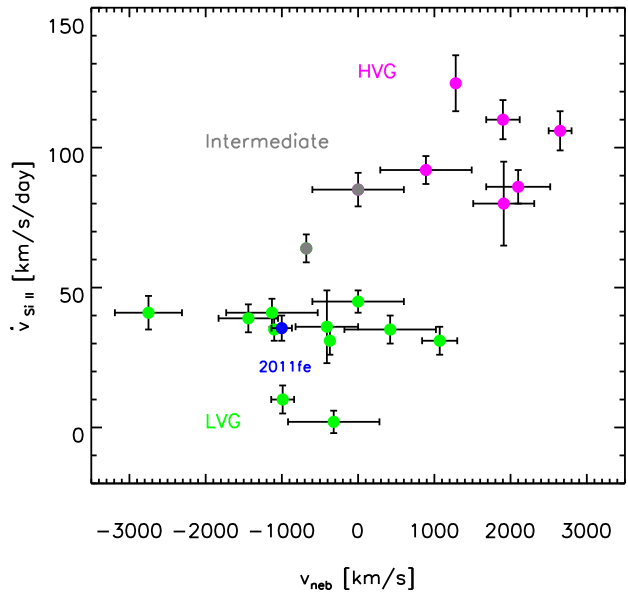}
\caption{\label{f:vdot_vs_vneb} Comparison of nebular velocity with respect to host galaxy rest frame to the early-time velocity gradient of \ion{Si}{2} (see \citealt{maeda10b} for details).  SN~2011fe is labeled in blue amongst the green Low Velocity Gradient (LVG) SN.  The gray, ``Intermediate'' SN include faint events SN~1986G and SN~2007on, while the High Velocity Gradient (HVG) is colored pink.  Peculiar SN such as SN~2004dt have not been included in this plot.}
\end{figure}

%%%%%%%FIGURE VNEB
The large difference in velocity between the [\ion{Fe}{2}] and [\ion{Fe}{3}] features places SN~2011fe firmly in the ``low-velocity gradient'' group (see Figure~\ref{f:vdot_vs_vneb}).  We confirmed this by examining the nebular velocity (defined by \citealt{maeda10} as the average of the [\ion{Fe}{2}] 7155 \AA\ and the [\ion{Ni}{2}] 7378 \AA\ velocity shifts) and the early-time \ion{Si}{2}\ velocity gradient, which we calculated using the spectra presented by \citet{parrent12} and made publicly available through the Weizmann Interactive Supernova data REPository\footnote{http://www.weizmann.ac.il/astrophysics/wiserep/}.  
The observed velocity shift of the singly ionized nebular lines implies a viewing angle
of roughly 70$^\circ$ from the symmetry axis as defined by \citet{maeda10b}.

\section{Conclusions}

We have analyzed mid-IR photometry and optical spectroscopy of the nearby type~Ia supernova
2011fe at ages between six months to a year after the explosion. We found the following:

\begin{itemize}

\item The luminosity in the Spitzer/IRAC bands are well fit by an exponential fading. The
observed decay rate was 1.48$\pm 0.02$ \mag100\ in the 3.6~$\mu$m band and 0.78$\pm 0.02$ \mag100\
at 4.5~$\mu$m. The fading rate in the CH1 filter is similar to that seen for many SNIa
in the optical $B$ and $V$ filters. The CH2 decline rate is slower than the $^{56}$Co radioactive
decay rate.

\item Using SN~2011fe and three other SNIa observed by Sptizer, we found that the
ratio of the flux at 3.6~$\mu$m to the flux at 4.5~$\mu$m at an age of 230 days
correlates with the light curve decline rate at maximum brightness. Since the
light curve shape at maximum is also correlated with the peak luminosity,
we expect the CH1$-$CH2 color index at a fixed age is a good indicator of the
radioactive nickel yield of a SNIa.

\item The series of optical spectra obtained with the LBT show emission features
fading at three different rates depending on the element and ionization state. Singly ionized
iron-peak elements faded close to the $^{56}$Co radioactive decay rate. Doubly ionized iron
lines decayed at a rate similar the $B$-band fading. Finally, doubly ionized cobalt faded at
more than twice the $^{56}$Co radioactive decay rate.

\item The optical and near-IR spectra of SN~2011fe show large ($> 1000$ km$\;$s$^{-1}$)
velocity shifts in the singly ionized emission lines while doubly ionized lines showed significantly
smaller offsets. 

\end{itemize}

It has long been known that the decline in radioactive energy
input is the cause of the fading luminosity in the nebular phase. Here, we show that the doubly
ionized species fade more quickly than the radioactive energy input and suggest that this
additional fading is due to recombination. Additionally, doubly ionized cobalt sees a decline in
excitation, recombination losses and direct radioactive decay to iron and so it fades at the fastest
rate of all the emission lines. The decay rate variation with element and ionization state demonstrated in the LBT spectra
nicely explains the range of observed decay rates seen in broadband filters during late-time
studies of SNIa \citep{lair06}.

Our mid-IR Spitzer photometry is consistent with the slow fading of singly ionized iron in the CH2 band and
doubly ionized iron or cobalt in the CH1 filter. From the correlation between the IRAC color index with
the radioactive nickel yield, we argue that \ion{Co}{3} is probably the primary source of emission around
3.6~$\mu$m up to a year after explosion. The rapid fading of the \ion{Co}{3} emission will eventually lead
to \ion{Fe}{3} or \ion{Fe}{2} becoming dominant in CH1 and a decelerating decay rate in the future.

We also measured the individual line velocities of optical and NIR features and confirm the pattern of
velocity offsets seen by \citet{maeda10}. That is, emissions from singly ionized species have higher
velocity offsets than those from doubly ionized elements. According to \citet{maeda10b}, this is because the singly
ionized species reside in the dense inner region of the nebula where an offset deflagration occurred, while
the doubly ionized atoms occupy the outer, low density and more symmetric regions of the nebula.
Both the measures of the early and late spectra indicate SN~2011fe was a typical `low velocity gradient' SNIa
viewed at an angle of about 70$^\circ$ relative to the offset axis of the deflagration.

\acknowledgements
{\it Acknowledgements}
We thank L. Storrie-Lombardi for help in scheduling
the Spitzer observations.  We acknowledge partial funding
of this research through the Spitzer guest observer program
(PID 80196).

The MODS spectrographs were built with funding from the NSF grant
AST-9987045 and the NSF Telescope System Instrumentation Program
(TSIP), with additional funds from the Ohio Board of Regents and the
Ohio State University Office of Research.

The LBT is an international collaboration among institutions in the United
States,
Italy and Germany. LBT Corporation partners are: The University of Arizona
on behalf
of the Arizona university system; Istituto Nazionale di Astrofisica,
Italy; LBT
Beteiligungsgesellschaft, Germany, representing the Max-Planck Society,
the
Astrophysical Institute Potsdam, and Heidelberg University; The Ohio State
University,
and The Research Corporation, on behalf of The University of Notre Dame,
University of Minnesota and University of Virginia.

%%%%%%%%%%%%%%
%	BIB
%%%%%%%%%%%%%%

%%%%%%%%%%%%%%
%	TABLES
%%%%%%%%%%%%%%

\clearpage
\begin{deluxetable}{cccc}
\tabletypesize{\scriptsize}
\tablecaption{\label{t:photometry} IRAC photometry for selected Type~Ia Supernovae}
\tablewidth{0pt}
\tablehead{
\colhead{SN}	&\colhead{phase [days]}	& \colhead{CH1 Flux [$\mu$Jy]}	& \colhead{CH2 Flux}
}
\startdata
2008Q     &	222	&	$10.21\pm1.08$&	$8.96\pm1.51$\\
2008Q     &	560	&	$0.38\pm0.28$&	...\\
2008Q     &	739	&	$<0.42$&	...\\
2009ig	&       196       &		$13.25\pm      1.41$&      $2.44\pm     0.69$\\
2009ig	&       375	      &	$2.87\pm      1.05$&      $1.53\pm     0.81$\\
2011fe	&       146       &       $656.73  \pm     13.60$&	$157.57\pm       6.92$\\
2011fe	&       167       &       $483.12  \pm     11.91$&	$120.97\pm       6.02$\\
2011fe	&       234       &       $197.35  \pm     7.99$&	$84.87\pm       5.14$\\     
2011fe	&       351       &       $40.89  \pm     4.25$&	$34.28\pm       3.40$\\
\enddata
\end{deluxetable}

\begin{deluxetable}{ccccc}
\tabletypesize{\scriptsize}
\tablecaption{\label{t:irslopes} Decline rates from least-squares fits to SN~Ia IRAC photometry, and IR color differences.}
\tablewidth{0pt}
\tablehead{
\colhead{SN}	& \colhead{$\Delta m_{15}(B)$} & \colhead{$\Delta$ CH1}	& \colhead{$\Delta$ CH2} & \colhead{CH1-CH2}  \\
\colhead{ }       & \colhead{(mag)} & \colhead{ [mag/100$^d$]}             & \colhead{[mag/100$^d$]} & \colhead{@230 days [mag]}
}
\startdata
2005df	& $1.09\pm0.04$	&	$0.97\pm0.03$			&	$0.65\pm0.09$  &	$-0.64\pm0.28$	\\
2008Q      & $1.36\pm0.06$	&     $1.06\pm0.24$	                &     ...                         &	$0.34\pm0.22$	 \\
2009ig	& $0.89\pm0.02$	&	$0.93\pm0.23$			&	$0.28\pm0.36$	  &	$-1.14\pm0.41$	\\
2011fe	& $1.07\pm0.06$	&	$1.48\pm0.02$			&	$0.78\pm0.02$	  &	$-0.44\pm0.08$	\\
\enddata
\end{deluxetable}

\begin{deluxetable}{clcccc}
\tabletypesize{\scriptsize}
\tablecaption{\label{t:lineshift} Nebular line identifications, velocities and decay rates.}
\tablewidth{0pt}
\tablehead{ \colhead{} &
\colhead{Identification\tablenotemark{a}}	& \colhead{ID Reference\tablenotemark{b}}	& \colhead{Average Wavelength\tablenotemark{a}} &	\colhead{$v_\mathrm{neb}$}	&	\colhead{Decline Rate} \\
\colhead{} & \colhead{} & \colhead{} & \colhead{} & \colhead{(km$\;$s$^{-1}$)\tablenotemark{c}} & \colhead{(\mag100)}
}
\startdata
 & [Fe~III] 4658 + [Fe~III] 4701 + [Fe~III] 4734	& M10 &	4698 &	$-457\pm$89	&	1.41		\\
 & [Fe~III] 4990 + [Fe~II] 5006				& B97 &	4998 &	$-1638\pm$180 &	1.39		\\
 & [Fe~II] 5262 + [Fe~III] 5270				& M10,B97 & 5266&	$-962\pm$296 &      1.55		\\
 & [Co~III] 5890 + [Co~III] 5908				& B97 & 5899 &	$-422\pm$56 &		2.01		\\
 & [Co~III] 6129 + [Co~III] 6197				& B97 & 6163 &	$+146\pm$165 &     	2.33		\\
 & [Co~III] 6578 + [Fe~II] 6456 + [Fe~II] 6516	& B97,NIST & 6517 &	$+1391\pm$1850\tablenotemark{d}	& 1.65	\\
 & [Fe~II] 7155 + [Fe~II] 7172				& M10,B97 & 7154 &	$-1006\pm$59	&  	0.85		\\
 & [Ni~II] 7378 + [Fe~II] 7388				& M10,B97 & 7383 &	$-1137\pm$102 &    0.81		\\
 & [Fe~II] 8617							& M10 &	8617 &	$-1559\pm$101 &     0.47		\\
 & [Fe~II] 1.257$\mu$m					& M10 &	1.257$\mu$m & 	$-1223\pm$95\tablenotemark{e}  &	...	\\
 & [Fe~II] 1.644$\mu$m					& M10 &	1.644$\mu$m & 	$-1137\pm$197\tablenotemark{e}	& ...	\\	
\enddata
\tablenotetext{a}{Wavelengths are given in \AA\ unless otherwise noted.}
\tablenotetext{b}{References: (M10) \citealt{maeda10}; (B97) \citealt{bowers97}; (NIST) \citealt{kramida12}}
\tablenotetext{c}{Velocities are corrected for the M101 redshift of $z=0.000804$ listed by NED (NASA/IPAC Extragalactic Database; http://ned.ipac.caltech.edu)}
\tablenotetext{d}{The large uncertainty derives from an assumed 40 \AA\ error in fitting the highly-blended line profile.}
\tablenotetext{e}{from LBT/LUCIFER spectrum obtained at age 234 days.}
\end{deluxetable}

\end{document}